\documentclass[11pt,twoside]{article}
\usepackage{asp2006}

\usepackage{epsf}
\usepackage{psfig}
\usepackage{lscape}

\begin{document}
\title{Star formation and structure formation in galaxy interactions and mergers}   %%% Fill in title
\author{Frederic Bournaud}   %%% Fill in author names
\affil{CEA Saclay, IRFU, SAp, 91190 Gif-sur-Yvette, France. {\tt frederic.bournaud@cea.fr} }    %%% Fill in author affiliations

\begin{abstract} 
A number of theoretical and simulation results on star and structure formation in galaxy interactions and mergers is reviewed, and recent hydrodynamic simulations are presented. The role of gravity torques and ISM turbulence in galaxy interactions, in addition to the tidal field, is highlighted. Interactions can drive gas inflows towards the central kpc and trigger a central starburst, the intensity and statistical properties of which are discussed. A kinematically decoupled core and a supermassive central black hole can be fueled. Outside of the central kpc, many structures can form inside tidal tails, collisional ring, bridges, including super star clusters and tidal dwarf galaxies. The formation of super star clusters in galaxy mergers can now be directly resolved in hydrodynamic simulations. Their formation mechanisms and long-term evolution are reviewed, and the connection with present-day early-type galaxies is discussed.
\medskip

\noindent
This review is an extended version of the proceedings of the ``Galaxy Wars: Stellar Populations and Star Formation in Interacting Galaxies Conference'', B. Smith, N. Bastian, J. Higdon and S. Higdon Eds., to be published in the ASP Conference Series.
\end{abstract}

\section{Introduction}
Galaxy interactions and mergers drive gas response and subsequent star formation. A large variety of stellar structures can then be formed : decoupled cores, super star cluster, tidal dwarf galaxies, etc.. I review results from numerical simulations related to the formation of stars, star clusters and other structures in galaxy collisions. Section~2 reviews some general processes that drive gas dynamics and star formation in galaxy interactions: the tidal field, gravity torques, increased turbulence. Section~3 relates to star formation in the central kpc, in particular the intensity of merger-driven starbursts, and their consequences like the formation of kinematically decoupled cores. In section~4, I review different processes that take place mostly outside of the central kpc: star formation in tidal tails and collisional rings, formation of super star clusters (SSCs) and tidal dwarf galaxies (TDGs). Section~5 briefly discusses the long term survival of the newly formed structures around merger remnants and the relation to present-day early-type galaxies.

\section{Theoretical framework: gas response in interactions and mergers}
\paragraph{The tidal field}

A first driver of the gas response in galaxy interaction is the tidal field. Typically, the tidal field of a single galaxy is disruptive at large radius and outside its own disk, but can be compressive in the inner regions if the mass distribution has a low Sersic index. During a distant interaction or the early stages of a merger, a given galaxy is then mostly affected by a disruptive tidal field from its distant companion. More advanced mergers can have compressive tidal field over large regions, when the galaxies begin to overlap \citep[e.g.,][for the Antennae]{renaud}. 

While a disruptive tidal field naturally tends to expell material from disk galaxies, it is not the only mechanism responsible for the formation of long tidal tails, and it cannot explain central gas inflows: this is in fact mostly driven by gravity torques.

\begin{figure}[!ht]
\centerline{
  \psfig{figure=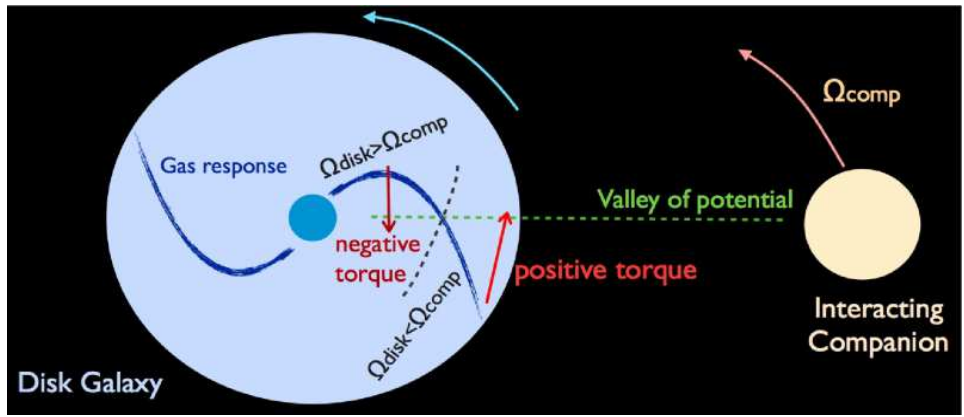,width=5in} }
\caption{Simplified description of the gas response in a galaxy interaction. Inside the corotation radius, gas concentrates on the leading side of the valley of potential and  undergoes negative gravity torques, which drives a central gas inflow. At the opposite, positive gravity torques dominate outside the corotation, which drives lot of the outer disk material in the so-called ``tidal'' tails.}
\label{fig:1}
\end{figure}

\paragraph{Gravity torques, inflows and outflows}
The tidal field from a companion breaks the symmetry of the gravitational potential. This induces a response of the disk material, in particular its cold gas, which is more pronounced for prograde\footnote{in which the orbit and disk rotations have similar orientations} interactions. The gas can form an interaction-driven pair of grand-design spiral arms, like modeled in detail for instance for M51 by Dobbs (2009, this meeting) -- note though that grand-design spiral do not require an interaction to form \citep[e.g.,][]{ET93}. The gas response is often more complex than a pair of spiral arms, but a general feature remains: inside the corotation radius\footnote{The corotation radius is the radius at which the rotation speed of the studied disk equals the orbital speed of the companion.}, the gas preferentially concentrates on the leading side of the valley of potential; outside of the corotation it concentrates on the trailing side. A similar but better-known gas response is found in barred galaxies, where the disk symmetry is broken about the same way by the barred pattern \citep{CG85,athanassoula}. A corotation does not always exist and can move with time, but is in general located at a few kpc of radius in the disk. The subsequent process is illustrated on Figure~\ref{fig:1}: inside the corotation, the gas undergoes negative gravity torques, and loose angular momentum in a rapid central gas inflow towards the central kpc or less. In the outer disk, gas would gain angular momentum and fly out to larger radii in long tidal tails. 

It is worth noting that:
\begin{itemize}
\item The so-called ''tidal'' tails do not just come directly from the tidal field. Gravity torques induced by the interaction are an important process in tidal tail formation. If the tidal field was the main factor, few mergers would actually have a central gas inflow and related central starburst. This was already illustrated by the restricted three-body simulations by \citet{toomre77}: the tidal field was included but not the full disk response, tidal tails were forming but relatively short and no central gas inflow was induced. The role of disk self-gravity and gravitational torques in amplifying the disk response was noted in \citet{toomre81}

\item Stars in a disk galaxy behave in a roughly similar way, but their much larger velocity dispersion and collisionless dynamics makes the gravity torques much weaker. This is why stars pre-existing to a galaxy interaction are generally not found in great amounts in tidal tails (e.g. Duc et al. 1997, but see for instance Duc et al. 2000 in Arp245). For the same reason, there is no significant central inflow of pre-existing stars (while the inflow of gas can form a lot of new stars in the center).
\end{itemize}
{\bf Collisional ring} galaxies form through a different dynamical process. The central impact of a companion onto the disk triggers large epicyclic motions around the initial orbits of stars and gas clouds. The superposition of motions and the variation of the epicyclic frequency with radius result in an expanding wave of material. This process is thoroughly reviewed by \citet{appleton}. The ring wave is initially an unbound feature, like the spiral waves in an isolated disk, but the ring can become self-gravitating, bound, when enough mass is gathered in the ring \citep[e.g.,][]{B07}. While the initial process differs, collisional rings and are relatively similar to tidal tails in terms of substructure formation, so they will hereafter be treated as the same kind of "tidal" structures.

%above done

\paragraph{Increased gas turbulence and shocks}

There are other physical processes that influence structure formation on small scales, while tidal tails and collisional rings are formed. For instance, strong shocks can be induced if two disks in an interacting galaxy pair directly collide in a coplanar merger or on some other specific orbit -- such specific situations can lead to shock-induced star and star cluster formation, which have been studied for instance by \citet{barnes} and \citet{saitoh}. This specific case will not be treated in the following parts, I refer the reader to the references above.

A more general and equally fundamental process that occurs in any galaxy interaction is the triggering of gas turbulence by interactions. This can be observed as an increase in the cold atomic or molecular gas velocity dispersion (linewidth). It has been observed and modeled in details over the last decade in particular by \citet{2163A,2163B,2163C} in the interacting pair NGC2136/IC2207, and is also observed in many other systems, for instance \citet{lisenfeld}  for the molecular gas in an interacting pair.
The origin of the increased turbulence is that gas from various regions of the disk moves along non-parallel orbits towards the tidal and collisions features. An important consequence is an increased Jeans mass, which is the typical mass of star-forming gas clouds in an unstable medium. Not only the gas complexes can be more massive, but also the star formation in these can be more efficient before gas expulsion, because of the increased pressure \footnote{The turbulent pressure increases since the velocity dispersion increases, and the thermal pressure can increase, too \citep{li}}. Star clusters can then form following the same universal process as in isolated galaxies, but with higher characteristic masses and efficiencies \citep{EE97, E02}. 

\medskip 

The capability of numerical simulations to reproduce the increased ISM turbulence in interacting galaxies is an important aspect for their ability to describe star formation in a realistic way. However, many numerical codes need to impose a temperature floor below which gas cooling is not resolved, quite often around $10^4$~K in {\em SPH simulations}. This implies that the thermal velocity dispersion is always high and the variation of the turbulent speed may not be resolved in such simulations. In this regard, {\em sticky particle models}\footnote{which is a dynamical model for a cold, star-forming interstellar medium, dominated by the turbulent pressure more than the thermal pressure} can be preferred to study star and star cluster formation, since they have no artificial thermal pressure floor in excess of the turbulent pressure \citep[see discussions in][]{semelin,martig}. For instance, the mass fraction of gas that lies in regions with a turbulent speed above 30~km~s$^{-1}$ is increased by a factor of 5 by the interaction, in the high-resolution major merger simulation by \citet{BDE08}.   A very high spatial resolution can also allow a low temperature floor and thermal sound speed in {\em grid hydrodynamic models} with adaptive resolution (AMR), around 100~K and 1~km~s$^{-1}$ in the best cases \citep[e.g.,][]{tasker}. Such an AMR code has been recently used for the first time for galaxy mergers by \citet{kim}. Their model overall reproduces a multiphase ISM reaching relatively low temperature of $\sim 100$~K in the densest regions, with a maximal resolution of 3.8~pc (for relatively small galaxies), but a lower effective resolution inside the tidal tails, inside which the largest structures are resolved, but the smaller substructures might be missed. Other AMR simulations by Chapon et al. (in preparation, see later and Fig.~5) confirm that adaptive grid methods can be successful in resolving the turbulent and cloudy structure of the ISM, and the formation of dense gas clouds and star clusters in interacting galaxies.

A critical aspect of AMR and adaptive methods in general is the presence of low-resolution regions. One aspect is the maximal spatial resolution, but another aspect is the refinement strategy and the mass resolution used to refine the adaptive grid. Depending on this, the spatial resolution can be very high with a very low temperature floor in the densest central regions, but a large fraction of the gas could lie in regions where the spatial resolution is not so high, preventing the formation of a cold and turbulent phase from being resolved. A high {\em mass } resolution is as important as a high spatial resolution to resolve the formation of stellar structures everywhere in a merging pair of galaxies.

\medskip

To summarize, tidal fields, gravity torques leading to central inflows and outer outflows, increased turbulence, high Jeans masses and high star formation efficiencies are the key ingredients for the formation of stellar structures in galaxy mergers. Some other processes like extended shocks can also occur on specific collision orbits.

\medskip

All these mechanisms are most efficient in major mergers, with mass ratios from 1:1 to about 3:1 -- the mergers that would directly transform a pair of spiral galaxies into an elliptical-like spheroid. In minor mergers (for instance a 10:1 mass ratio), the same interaction-induced processes happen, but they are weaker for the largest galaxy. They can strongly affect the low-mass companion, but this represents only a small fraction of the total mass. Structure formation in minor mergers has not been extensively studied in numerical simulations. In observations, it seems that contrary to major mergers, minor mergers can have most star formation in relatively undisturbed gas disks \citep[e.g.,][]{trancho}, and are relatively poor in tidal tails and other collisional structures compared to major mergers \citep[e.g.,][]{manthey}. The results reviewed in the next parts are mostly focused on major mergers, which include both mergers of equal-mass galaxies (1:1) and mergers of unequal- but sizeable-mass galaxies (mass ratios around 2:1 to 4:1).

%above done

%%%%%%%%%%%%%%%%%%%

\section{Inner structures: central starbursts and kinematically decoupled cores}

\paragraph{Merger-induced starbursts}
One of the main effects of a distant interaction or early stage merger is the central gas inflow. It increases the gas density, the gas clouds collision rate, so that any model for star formation (density-based models, e.g. Elmegreen 2002, or collision-induced star formation, e.g. Tan 2000) will precict a significant increase in the star formation rate, or a so-called starburst. This has been extensively studied in simulations since the early 90s, in relation with UltraLuminous Infrared Galaxies and other kinds of starbursting galaxies.

The first simulations of mergers including gas dynamics have shown that very strong starbursts can be triggered by major mergers: compared to the same galaxies in isolation, the star formation rate can increase by factors of tens, or even sometimes hundreds (these values refer to the peak of the starburst, not the average SFR over the whole duration of the merger). This is shown for instance in \citet{mihos, cox1, springelh, robertson2, hopkins}. 
The starburst usually peaks in the early phases of the merger, near the first pericenter passage of a galaxy pair. A secondary but generally weaker burst can be triggered during the final coalescence. The efficiency of merger-induced starbursts rapidly drops with increasing mass ratios \citep[i.e. more minor mergers,][]{cox2}. 
	
Early studies are likely biased towards the most favorable cases (prograde resonant orbits, coplanar disks, etc..). They reproduce the mechanism, but are not statistically representative of random mergers. A very extensive, public\footnote{available at: {\tt http://galmer.obspm.fr}} library comprising hundreds of galaxy mergers with various orbits, mass ratios and galaxy types, called {\sc ``Galmer''}, has been produced by Combes and collaborators \citep[see][]{galmer-paper,chilingarian}. Using this library, \citep{dimatteo1,dimatteo2} have shown that major mergers generally trigger only {\em modest starbursts } \citep[see also][for the influence of an additional larger-scale, group or cluster, tidal field]{martigbournaud}. On average, the peak star formation rate is 3--4 times that of the galaxy pair before the interaction/merger. It ranges from cases with almost no enhancement of the SFR to very efficient starbursts. Only about 10\% of major merger and interactions have a ``strong'' starburst if one defines this as a peak SFR at least ten times higher than in the pre-interaction galaxy pair. Statistics on the SFR are relatively similar for mergers and distant interactions, and do not depend much on the initial bulge mass and gas fraction of the interacting galaxies, so should not depend much on their redshift either. These robust results were confirmed comparing SPH simulations and sticky particle simulations, and using several star formation models (density-based Schmidt law, cloud-cloud collision rate, etc). Note that these numerical results are in good agreement with recent observations pointing out that interactions and mergers drive only a small fraction of star formation galaxies at various redshifts \citep[see for instance][]{jogee,robaina}.

While mergers generally experience only a moderate starburst, this remains an important phenomenon because:
\begin{itemize}
\item This star formation activity often takes place in dense and massive clusters (see Section~4.1).
\item The typical duration of merger-triggered starbursts is $\sim 0.5$~Gyr (for the masses of bright spiral galaxies, not dwarfs -- \citet{dimatteo2}). A pair of Milky-Way like galaxies merging on random orbits would then, on average, form a few $10^9$~M$_{\sun}$ of new stars from merger-induced processes, which is not a negligible mass.
\item While there is relatively extended star formation in tidal tails, a large fraction ($\sim$ 50\%) of the new stars generally form in the central kpc: the densities of the new stars in the central regions can be quite high compared to the density of the pre-existing stars. This contributes to a general increase in the concentration and Sersic index of major and minor merger remnants \citep{naab1,naab2,bournaud07}. The high central density of the newly formed stars can also form decoupled central structures as KDCs.
\end{itemize}

\paragraph{Formation of KDCs}

\begin{figure}[!ht]
\centerline{
  \psfig{figure=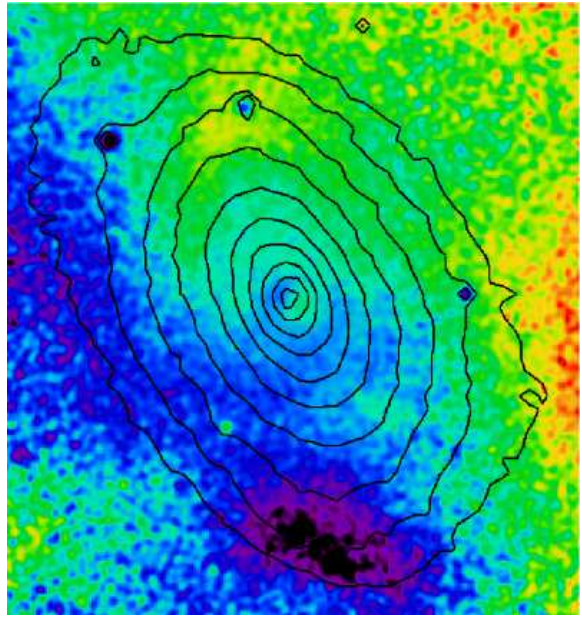,width=3in} }
\caption{Stellar velocity field of a major merger remnant (from Bournaud et al. 2008 and Bois et al. 2009). The image size is 10~kpc, the black contours represent the projected stellar density (log spacing) and the colors code the line-of-sight velocity field. A $\sim$500pc-large KDC is visible at the center of this elliptical-like galaxy. It results from the merger-driven gas inflow and nuclear star formation.}
\label{fig:kdc}
\end{figure}

The interaction-driven gas inflow and central burst of star formation can result in the formation of a kinematically decoupled core (KDC). The inflowing gas has been torqued inwards from a certain portion of the two disks, so generally gets misaligned with the total spin of the interacting galaxy pair (or the spin axis of the resulting elliptical galaxy once the merger is relaxed). Star formation in this central gas disk then results in a misaligned young stellar system in the central regions, typically 100~to~1000~pc in radius, which is a newly-formed KDC. An example of such a KDC formed in a wet major merger simulation is shown on Fig.~\ref{fig:kdc}. Another example of KDC formation in early-type galaxy can be found in \citet{naab-kdc}. A prototype galaxy harboring such a KDC resembling the result of major merger can be for instance NGC~4365, the KDC of which is well resolved in the Sauron stellar velocity field \citep{ems4365}. For more details on the general orbital structure of merger remnants, I refer the reader to \citep{jesseit} and references therein.

More extended kinematically decoupled or counter-rotating structures (but maybe not "cores") can also be formed by the merging and superimposition of two disks initially spinning in opposite directions, or by the accretion of external gas, like for instance in NGC~4550 \citep{puerari,afc4550,afc2768}, or in Elliptical+Spiral mergers \citep{dimatteokdc}. These decoupled structures are generally more extended than the central, compact KDCs formed by the merger-induced gas inflow and star formation as described above.

\paragraph{Black Hole growth}

On the very central scales, the main consequence of galaxy interactions and mergers can by the growth of black holes. The interaction-driven gas inflows, that drive the central starburst, can fuel a central super massive central black hole (SMBH) \citep[e.g.,][]{combes03}. When two merging galaxies host SMBHs, the two SMBHs may eventually merge together. This can keep the merger end-product in agreement with the observed black hole mass -- stellar velocity dispersion (or stellar spehroid mass) relation \citep{johansson, tdimatteo,robertson}. An important issue, though, is that the velocity kick resulting from the merging of two black holes can expulse the resulting SMBH from the center of the galaxy, or even expulse it from the whole galaxy potential well \citep[e.g.,][]{kick1,kick2}. This may be avoided if the gas fraction is high enough for a nuclear disk to torque the SMBHs and align their individual spins before they merge with each other, which minimizes the velocity kick \citep{bogdanovic}. However, the general fate of merging central SMBHS remains unclear as fastly recoiling black holes are possibly observed \citep{recoil1,recoil2}.

%above done

%%%%%%%%%%%%%%%%%%%

\section{External structures: tidal tails, tidal dwarf galaxies, and massive star clusters}

\paragraph{Tidal tails, bridges, collisional rings}

The tidal field and gravity torques generally lead to the formation of tidal tails. Other types of features can form in interactions, too, such as (1) collisional rings, in head-on collisions, these expanding ring-shaped waves result from the rapid change in the deepth of the potential well \citep{appleton, bournaud07}, (2) accretion tails, where gas from one galaxy is captured by the other galaxy, possibly forming a polar ring \citep{BC03, smith08, hancock09}, (3) bridges, as for instance in the "Taffy" galaxies \citep{taffy}, etc.

In the following, I mainly discuss the case of classical tidal tails. Star formation and substructure formation in other features like rings is relatively similar: the gas can reach high densities, velocity dispersions and pressure in all these kinds of interaction-induced structures.

\subsection{Substructures in tidal tails: two formation mechanisms}
\paragraph{Accumulation of material in massive objects}

While the formation of bound gas clouds and new stellar structures (star clusters) in isolated galaxy disks usually results from local gravitational instabilities, a different mechanism can participate in tidal interactions. Some disk material can pile-up in some particular region and become bound, star-forming. This was first proposed using simulations by \citet{E93}: a dense region in a disk (proto-cloud) can be moved outwards as part of a tidal tail. There it can remain bound instead of fragmenting into an unbound complex of several smaller pieces, because of the increased velocity dispersion. This allows the formation of bound star-forming clouds of $\sim 10^8$~M$_{\sun}$ in tidal tails, while the isolated parent disk would harbor GMCs only up to $\sim 10^6$~M$_{\sun}$. A generalization of this pile-up process was proposed by \citet{duc04}, where new N-body simulations show that large amounts of material can be piled-up without needing a pre-existing proto-cloud in the initial disk. A large region of the outer disk or even most of it can pile up in a relatively small (1-2 kpc) region of a tidal tail, most often near the tip of the tidal tail. There, it can be dense enough, and the shear and tidal field can be low enough, for this massive accumulation of gas to collapse or remain bound, and form stars actively. 

The same "pile-up" process can occur in other structures can the classical tidal tails, e.g. in accretion tails or in bridges \citep{smith08, hancock09}.

%above done

This mechanism typically leads to the formation of very massive star forming structures ($\sim 10^{8-9}$~M$_{\sun}$ for Milky Way-sized parent galaxies). They are made-up mostly of gas and new stars formed locally. The pre-existing stars, present in the parent galaxies before the collision, generally have a too high random velocities, escape these newly-formed system, even though the lowest-velocity ones can remain bound and amount up to, say, one third of the total mass of the new objects. For the same reason, these new objects cannot capture dark matter from the halo of their progenitor galaxy as dark matter particles have much too high random dispersion velocities \citep[see also][and references therein]{bournaud09}. 

%above done

\paragraph{Gravitational instabilities in tidal structures}

The classical process of a local gravitational instability that happens in disk to form GMCs and star clusters can also take place in tidal tails: gas densities can be high, and stabilization by the differential rotation can be relatively low at large radius in tidal tails. Thus, tidal tails that are sufficiently dense can be gravitationally unstable, just like portions of the gas disk in spiral galaxies. 

\begin{figure}[!ht]
\centerline{
  \psfig{figure=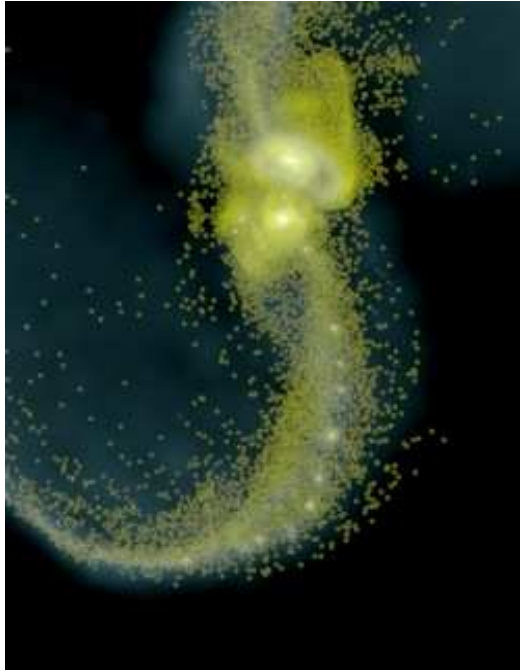,width=2.7in} }
\caption{SPH simulation of a major merger by \citet{wetzstein} showing massive star cluster formed by gravitational instabilities in a gas-rich tidal tail, in a ``beads on string''-like star formation mode. (Figure courtesy from Markus Wetzstein)}
\label{fig-w}
\end{figure}

There are differences though: (i) the higher velocity dispersion increases the typical Jeans Mass at which gas clouds form, (ii) the higher pressure can increase the star formation efficiency, and (iii) tidal tails usually have a high gas-to-star surface density ratio, which this ratio is low in modern spiral disks, so that the process is closer to a gas pure instability, with clouds forming at the gaseous Jeans mass, and separated by the gaseous Jeans length. Cloud formation and star formation in classical disks happens in a two-phase medium, where the gravity of gas and stars add \citep{jog}, and instabilities grow at all scales from the gaseous Jeans length to the stellar Jeans length \citep{E95}. Gas cloud and star cluster formation in tidal tails is thus characterized by massive clouds/clusters, with a typical mass and separation between the star-forming knots, giving a typical "beads on string" appearance. This is well seen in simulations, for instance those by \citet{wetzstein} (Fig.~\ref{fig-w}), and is quite consistent with observations \citep[e.g.,][]{weilbacher, smith08}.

The structures formed by this local collapse mechanism can be quite massive ($\sim 10^{7}$~M$_{\sun}$ for the biggest ones) but not as massive as the ones formed by the pile-up mechanism described earlier. They can be found all along tidal features, not just at the tip of tails. They are relatively poor in old stars compared to their progenitor disk galaxies -- the fraction of old stars is sometimes as large as 30\% or at most 50\%, which remains low compared to the $\sim$90\% fraction for typical $z=0$ disk. Also, they contain almost no dark matter from the halo of their parent galaxy, as their escape velocity is at most a few tens of km/s. They are also found frequently in collisional rings, when these are massive enough for instabilities to arise -- see \citet{struck2, horellou, B07} for simulations, and observed cases e.g. in Higdon (1995), Higdon et al. (1997), Pellerin (this meeting).

%above done

\paragraph{Two formation mechanisms and two types of substructures -- Tidal Dwarf Galaxies and Super Star Clusters?}

As described above, there are two main mechanisms for the formation of dense, gas-rich, star-forming substructures in galaxy interactions: (1) local gravitational instabilities in gas-rich tidal tails, rings, and bridges, and (2) the pile-up of large amounts of gas in particular regions.

Some merger simulations show these two mechanisms happening at the same time. This include the highest resolution model of a major merger (with gas) that has been performed by \citealt{BDE08} with a stick particle scheme, as well as a recent simulation by \citet{chapon} using a grid hydrodynamic code (AMR). These simulations are shown on Figure~4 and 5, respectively.

The objects produced by the pile-up mechanism, most often near the tip of tidal tails, are very massive ($10^{8-9}$), extended ($\sim$1~kpc), rotating disks ($V/\sigma$$>$1). A single merger would rarely form more than a couple of such objects. The objects produced by local gravitational instabilities can be much more numerous, sometimes forming "beads-on-string" clumps along tidal tails. They do not differ from the other category only by having lower masses (up to $10^{6-7}$~M$_{\sun}$), but also by being much more compact (10-100~pc even for the most massive ones), and supported by random stellar motions ($V/\sigma$$<$1). 

Because of their different formation mechanism and properties, these structures should not be considered as the high- and low-mass ends of the same family of objects. I actually propose that the pile-up mechanism forms massive Tidal Dwarf Galaxies (TDGs), which possibly survive as dwarf satellites around the merger remnant, and that gravitational instabilities form Super Star Clusters (SSCs). Even though the most massive SSCs could survive and resemble dwarf galaxies of tidal origin, this classification is probably more relevant given the differences reviewed above.

Observations suggest that SSCs along tidal tails are less numerous when massive TDGs are found at the tip of tidal tails \citep{knierman}. A simple explanation could be that the pile-up of gas into a massive TDG decreases the available mass for SSC formation, but this remains to be directly tested in simulations.

\begin{figure}[!ht]
\centerline{
  \psfig{figure=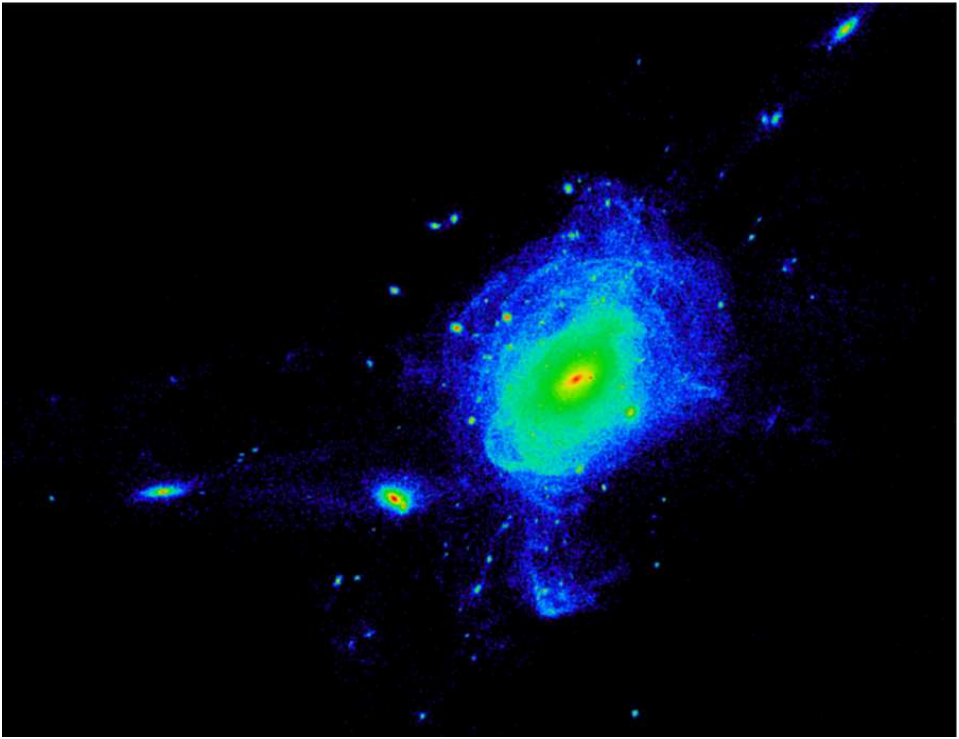,width=5in} }
\caption{Result of a gas-rich major merger \citep[from][]{BDE08}, with only the ``young'' stars formed during/after the merger shown. Three large, massive, rotating TDGs formed by pile-up of gas at the tip of tidal features (marked), with masses of a few $10^{8}$~M$_{\sun}$. Tens of compact SSCs with masses from $\sim$$10^{5}$~M$_{\sun}$ to $10^{7}$~M$_{\sun}$ formed by gravitational instability in the turbulent gas of tails, arcs, shells, and get randomized in the late stages of the merger. Nuclear star formation triggered by the interaction has formed a central KDC seen under a different projection on Figure~1.}
\label{fig:2}
\end{figure}

\begin{figure}[!ht]
\centerline{
  \psfig{figure=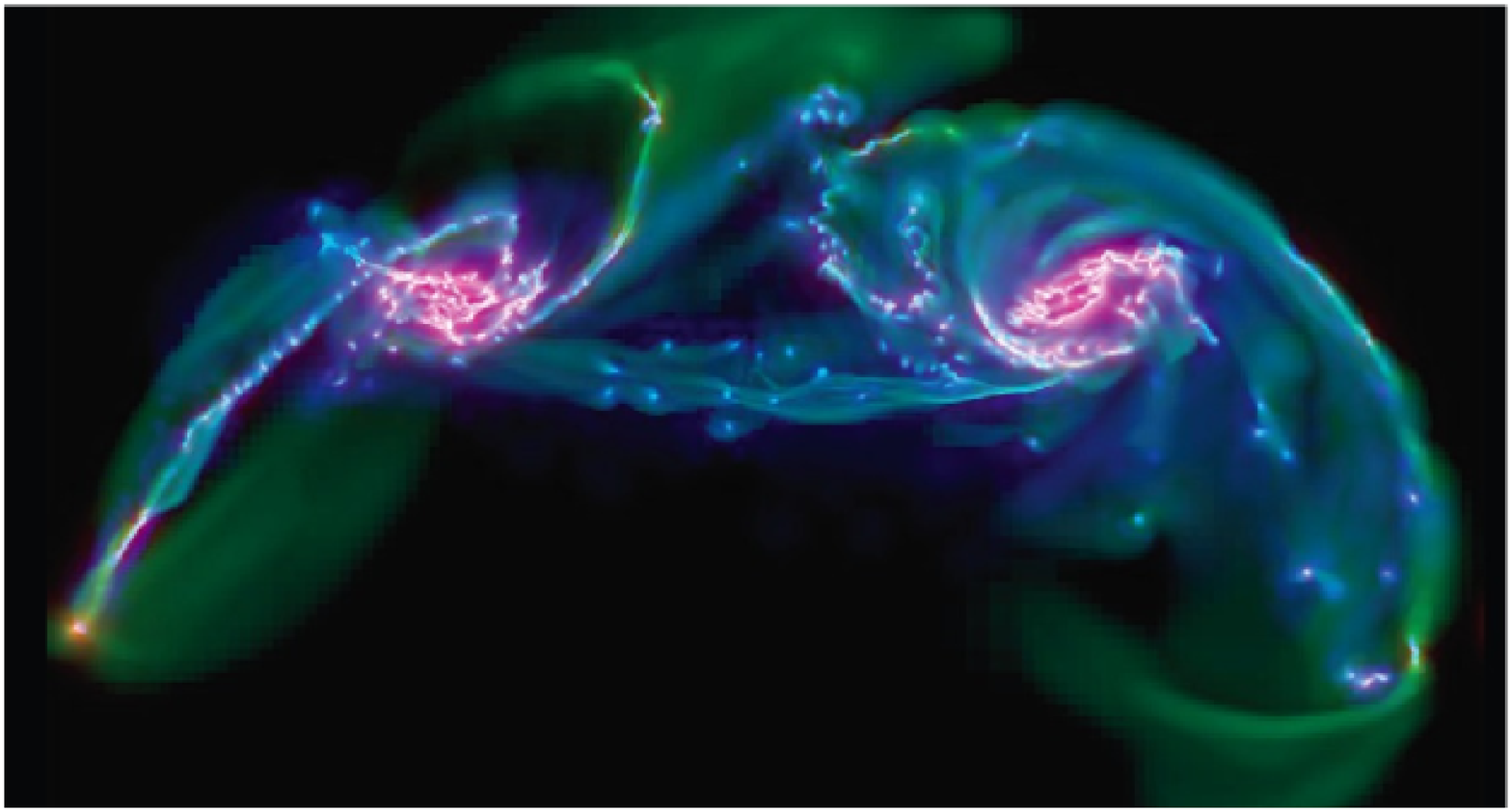,width=5in} }\vspace{.05cm}
\centerline{
  \psfig{figure=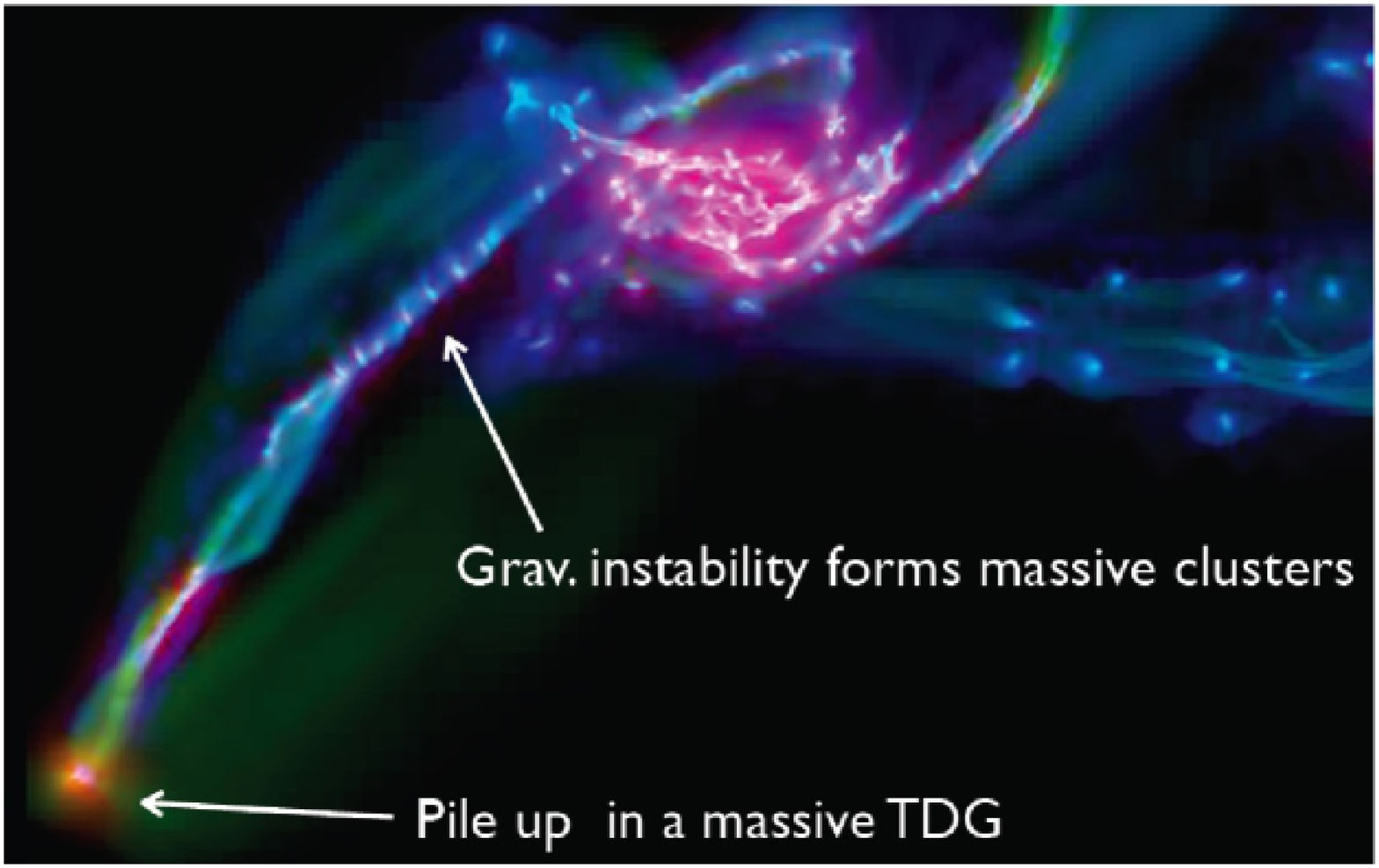,width=5in} }
\caption{Grid hydrodynamic (AMR) simulation of a major merger by \citet{chapon}. Pile-up of material in a massive TDG at the tip of a tidal tail is visible. Many massive SSCs form in the central arms and arcs and all along tidal tails, sometimes in a ``beads-on-string'' star formation mode.}
\label{fig:3}
\end{figure}

\subsection{Evolution and survival}
\paragraph{Long-lived TDGs?}

Massive TDG progenitors that form when material piles up in tidal tail might be disrupted by star formation feedback. Nevertheless, many of them form around or above the critical mass for dwarf galaxies to survive their initial starburst \citep{dekel-silk}. Their initial starburst is relatively weak, too: simulations show only moderate gas surface densities, since these objects are extended rotating disks, and star formation rates of at most 1-10~M$_{\odot}$~yr$^{-1}$, consistent with observations \citep[e.g.,][]{boquien}. Thus star formation feedback does not seem to be a major limitation to the survival of massive TDGs.

Next, TDGs could be disrupted by the tidal field of their own progenitor galaxy, or could fall back and be re-absorbed by their progenitor -- most of the tidal material falls back at low radii within about one billion year \citep{mihoshibbard}. Nevertheless, the preferential formation of TDGs in the outer regions of tidal tails maximizes the timescale of these effects. A large simulation sample in \citet{BD06} shows that a significant fraction of TDGs can survive several Gyrs, orbitating as satellite galaxies around the merger remnant -- which is usually transformed into an early-type galaxy by the merger itself. They may thus contribute to the population of dwarf satellites, in particular around massive early-type galaxies and maybe some satellites of spirals too \citep{OT00, kroupa}, but they probably do not outnumber the classical dwarf galaxies of cosmological origin (see detailed statistics in Bournaud \& Duc 2006).

\paragraph{From SSCs to globular clusters?}

The numerous SSCs that form by gravitational instability in tidal tails may have a different evolution because of their low masses and high densities. These star clusters form with very high pressures and high gas densities \citep{li,BDE08,chapon}. The star formation efficiency is then expected to be very high, so that these SSCs may likely remain bound after the expulsion of gas by the first generations of supernovae, and evolve into globular clusters. The model in \citet{BDE08} produces more than one hundred such tightly-bound globular-like clusters; the ICMF remains uncertain because of the limited statistics but is consistent with a power law.

\citet{recchi} modeled the evolution of such young tidal substructures and also conclude that they can survive their initial starburst and remain bound. They also proposed that, if surviving, the most massive of these SSCs could resemble dwarf galaxies, adding to the population of TDGs described above even if they are less massive and formed differently.

%above ok

\subsection{Other outer structures}
Stellar shells and streams are other structures formed in galaxy mergers. Shells are density waves that can be purely stellar, hence form even in dry mergers \citep[e.g.,][]{dupraz}. Streams form mostly in the early stages of minor mergers, when dwarf companion galaxies are partially or fully disrupted by the tidal field, or as remnants from tidal tails in major mergers. \citep[e.g.,][]{johnston} A simulation showing many streams from minor mergers is shown on Figure~5 and the long-term survival of these structures is discussed in section~5.1. Tidal tails can be captured by companion galaxies, and form polar rings there \citep{BC03, hancock09}. Polar rings can also form in mergers with polar orbits, through a collisional ring-like mechanism \citep{bekki}.

\subsection{A comment on merger-driven starbursts}

Recent results on merger-driven starbursts were reviewed in Section~3. We have now seen that star formation in interacting galaxies is characterized by high gas dispersions and the formation of massive star clusters. The high-resolution simulations by \citet{BDE08} and \citet{chapon} have $\sim$50\% of their star formation in SSCs. Most simulations used to study merger-induced starbursts however do {\it not} resolve the small-scale structure of the ISM and the formation of gas clouds/star clusters. This is the case for the large statistical studies described in Section~3. The characteristics of merger-induced starbursts (how strong they are, how long they last, etc) thus remain relatively uncertain in most simulations when star forming complexes are not well resolved.

%%%%%%%%%%%%%%%%%%%

\section{From galaxy mergers to post-merger galaxies}
\subsection{Persisting stellar structures as probes of the role of mergers in galaxy formation}

We saw above that mergers can form a large variety of structures. I now briefly discuss the long-term survival of these structures and how they could probe past, relaxed mergers.

\begin{figure}[!ht] 
\centerline{  \psfig{figure=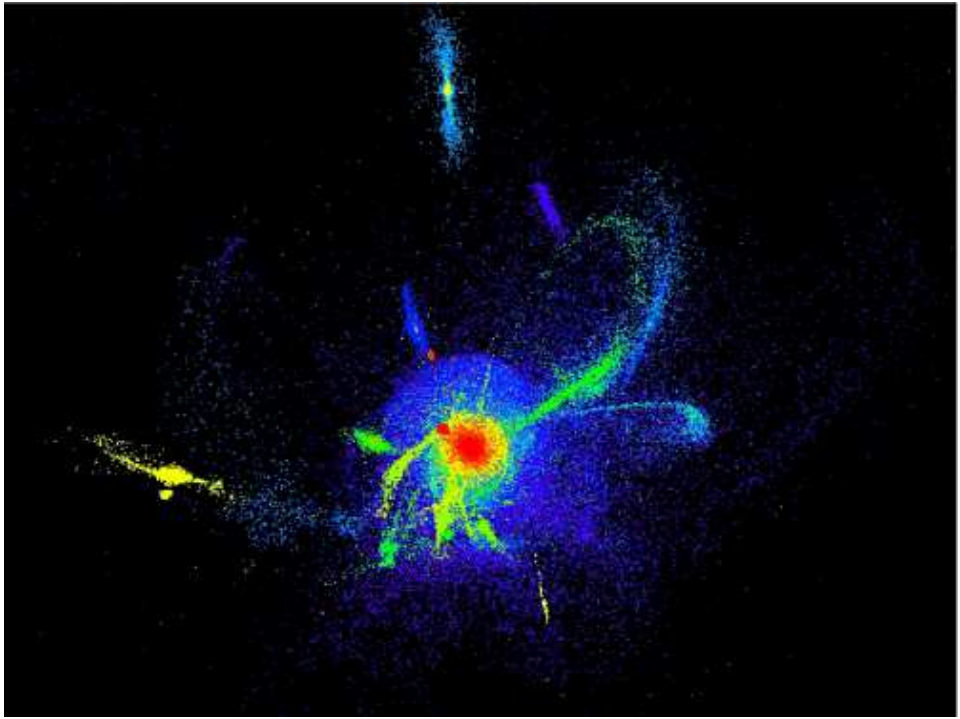,width=5in} }
\caption{Cosmological model by \citet{martig} of a $z=0$ elliptical galaxy, formed by several minor mergers at $z=1-2$ and a major merger at $z=0.2$. Many streams are visible around this galaxy, and were formed mostly during the minor mergers at $z > 1$, persisting after subsequent mergers. The colors code the stellar age, showing that each stream formed at a different epoch, in a different interaction. The figure shows about a 300$\times$400kpc area.}
\label{fig:4}
\end{figure}

\paragraph{Stellar streams}
The only known mechanism to form the observed streams is interactions and mergers, in particular with small companions. The streams can survive several billion years, hence tracing past mergers that do not have other obvious signatures. Interestingly, a cosmological zoom-in simulation by \citet{martig} on Fig.~\ref{fig:4} shows that streams formed in a first merger can survive more subsequent mergers, even a major one. The presence of a stream and its age likely trace the occurrence of a past merger each, and streams from interactions at $z>1$ may still be observable at $z=0$. The observation of streams can then be used to probe the role of hierarchical merging in galaxy assembly \citep[e.g.,][]{ibata, fardal}.

\paragraph{Globular clusters}
While most globular clusters (GCs) form at high redshift with a low metallicity, early-type galaxies have an additional population of higher metallicity GCs \citep[e.g.,][]{AZ92}. This is quite consistent with the simulation results above suggesting that the SSCs formed in galaxy mergers can survive and evolve into GCs. As GC formation requires violent processes to compress the gas into high-pressure regions, past mergers are the best-known explanation for the excess of high-metallicity GCs around early-type galaxies. 

\paragraph{KDCs} KDCs are another type of structure that can survive long after a merger and may provide evidence for past, relaxed major mergers. However KDCs may not be unambiguous evidence for past, wet major mergers as other mechanisms might also produce kinematically decoupled structures (see section~3).

\medskip

These were some examples of long-lived stellar structures that form in galaxy mergers and the observation of which may provide evidence for past mergers. As an example, the early-type galaxy NGC~4365 harbors both a compact KDC \citep{ems4365,davies} and numerous young, massive star clusters \citep{larsen4365}, which likely points towards this galaxy being the remnant of a past major gas-rich merger, in spite of NGC~4365 begin now a very relaxed, isolated galaxy.

%above ok

\subsection{Quenching of star formation: from starbursting mergers to "red and dead" early-type galaxies}

While mergers are an intense phase of star formation and structure formation, post-merger galaxies (i.e., S0s and ellipticals), are generally "red and dead" systems with very little star formation, if any. Gas consumption is cannot the key explanation in the termination of star formation: merger-induced starbursts are generally relatively weak \citep[][and section~3]{dimatteo2} and do not consume all of the available gas. Mergers happen mostly in dense environments, so a large fraction of today's red early-type galaxies are found in dense groups and clusters and have their cold gas stripped. As for those that are not in clusters, if they are massive enough to have the cold gas removed and further accretion stopped by AGNs and virial shocks \citep{cattaneo,birnboim1,birnboim2}. But there are also red spheroidal galaxies that are not in clusters, and below the critical mass for virial shocks and AGN heating \citep{weinmann} -- maybe not frequently at $z=0$, but likely more frequently at higher redshift. These cases may be best explain by not necessarily having their cold gas removed, but by the cold gas left-over after the merger and/or re-accreted later-on being stabilized by the massive stellar spheroid: this "morphological quenching" process is discussed in \citep{martig}. Indenpendantly of other environmental effects, this mechanism stabilizes the gas in any post-merger, early-type galaxy, ensuring that star formation becomes inefficient unless another interaction occurs or a high gas fraction is accreted. 

Thus the formation of stars and new stellar structures generally terminates after the intensely productive merger event. Star formation and structure formation could nevertheless resume after mergers if enough cold gas is accreted by merger remnants that do not lie in dense virialized clusters.

%above ok

%%%%%%%%%%%%%%%%%%%

\section{Conclusion}
Structure formation in galaxy interactions and mergers is quite rich. While mergers may not be the main driver of star formation in galaxies across the Hubble Time, they are a major process to explain the formation of long-lived small-scale stellar structures: massive bound (globular) clusters, tidal tails and streams, decoupled nuclear cores, etc.. Many of these structures can persist down to redshift zero as probes of the role of merger in the formation of early-type galaxies. We can expect deep observations and high-resolution simulations capable to trace low density structures do to be used in the future to better trace the structures left over by past mergers, and probe the rate, mass ratios, and other properties of these mergers.

\acknowledgements 
I am grateful to the organizers of this exciting meeting, and to Damien Chapon for providing me with his simulation results (Fig.~5) prior to publication. Collaborations and discussions on gas dynamics and star formation interacting galaxies with Pierre-Alain Duc, Elias Brinks, Fran\c{c}oise Combes, Bruce Elmegreen, Debbie Elmegreen, Eric Emsellem, Paola Di~Matteo and Romain Teyssier are gratefully acknowledged, as well as comments from Ji-Hoon Kim and Tom Abel. Support from Agence Nationale de la Recherche through grant ANR-08-BLAN-0274-01 is acknowledged.

\end{document}